\documentclass{elsarticle}
\usepackage{float}

\usepackage[english]{babel}
\usepackage{subcaption}
\usepackage{multirow}
\usepackage{tcolorbox}
\usepackage[hypcap=false]{caption}
\usepackage[]{hyperref}
\usepackage{booktabs}
\usepackage{enumitem}
\usepackage{pdfpages}
\usepackage{listings}
\newcommand{\new}[1]{\textcolor{black}{#1}}

\lstdefinestyle{promptstyle}{
  basicstyle=\ttfamily,
  frame=single,
  breaklines=true,
  commentstyle=\color{olive},
  morecomment=[l]{\#},
  breakindent=0pt,
  alsoletter={:.<>},
  keywordstyle=\color{orange}\bfseries,
  morekeywords={NetFlowSample,<...>,NF_Feature:Definition,IPV4_DST_ADDR,IpRelatedInformation, PreviousRelatedNetflows, PROTOCOL, L7_PROTO, Layer7ProtocolName, Layer4ProtocolUsedByProtocol, Layer7ProtocolDescription, Layer4ProtocolName, NetFlowSpecification, IndicatorsOfAttack, RelatedPreviousNetFlowConnections, eX-NIDSPromptTemplatePrefix, IPV4_SRC_ADDR,Layer7ProtocolName, Layer4ProtocolName, IpRelatedKnowledge},
}

\usepackage[a4paper, total={6in, 8in}]{geometry}

\begin{document}
\sloppy

\title{eX-NIDS: A Framework for Explainable Network Intrusion Detection Leveraging Large Language Models}

\author{Paul R. B. Houssel}
\author{Siamak Layeghy}
\author{Priyanka Singh}
\author{Marius Portmann}

\affiliation{organization={School of Information Technology and Electrical Engineering, University of Queensland},
            city={Brisbane},
            postcode={4072}, 
            state={QLD},
            country={Australia}}
\ead{p.houssel@uq.edu.au}
\ead{siamak.layeghy@uq.net.au}
\ead{priyanka.singh@uq.edu.au}
\ead{marius@ieee.org}

\begin{abstract}
%
This paper introduces \textit{eX-NIDS}, a framework designed to enhance interpretability in flow-based Network Intrusion Detection Systems (NIDS) by leveraging Large Language Models (LLMs).
In our proposed framework, flows labelled as malicious by NIDS are initially processed through a module called the \textit{Prompt Augmenter}. This module extracts contextual information and Cyber Threat Intelligence (CTI)-related knowledge from these flows. 
\new{This enriched, context-specific data is then integrated with an input prompt for an LLM, enabling it to generate detailed explanations and interpretations of why the flow was identified as malicious by NIDS.}
We compare the generated interpretations against a \textit{Basic-Prompt Explainer} baseline, which does not incorporate any contextual information into the LLM’s input prompt.
Our framework is quantitatively evaluated using the \textit{Llama 3} and \textit{GPT-4} models, employing a novel evaluation method tailored for natural language explanations, focusing on their correctness and consistency.
The results demonstrate that augmented LLMs can produce accurate and consistent explanations, serving as valuable complementary tools in NIDS to explain the classification of malicious flows. The use of augmented prompts enhances performance by over 20\% compared to the \textit{Basic-Prompt Explainer}.
\end{abstract}
\maketitle

\section{Introduction}
Large Language Models (LLMs) have revolutionised Natural Language Processing (NLP), excelling in tasks involving unstructured data such as text generation, contextual understanding, and language translation. Their great performance has led to widespread adoption in conversational AI, code generation, and beyond. However, applying LLMs in cybersecurity, specifically for Network Intrusion Detection Systems (NIDS), remains under-explored.

NIDS are essential for monitoring and analysing network traffic to identify malicious activities and potential security breaches. Current systems rely on a mix of signature-based methods, which detect known attack patterns, and anomaly-based techniques, which identify deviations from typical network behaviour. While deep learning-based NIDS have shown near-perfect performance on benchmark datasets~\cite{ahmad_network_2021, layeghy_benchmarking_2024}, they often suffer from a lack of explainability~\cite{neupane_explainable_2022}. This makes it difficult for security analysts to interpret, trust, and act on the detected threats, highlighting this need. Traditional explainable AI techniques which compute an importance score for each feature of the feature space, like SHapley Additive exPlanations (SHAP)~\cite{lundberg_unified_2017} have limitations. They require a strong understanding of machine learning, focus only on statistical anomalies, and lack contextual insights or external knowledge to explain feature importance. A limitation that potentially could be handled by LLMs. Although a few studies have explored using transformers and LLMs for threat detection~\cite{manocchio_flowtransformer_2024}, most adapt these models by replacing their sequence-to-sequence output with classification heads. While suitable for classification tasks, this modification eliminates the models' ability to provide explanations alongside their predictions, a key advantage of LLMs. As such, the potential of LLMs to improve explainability in NIDS remains underutilised. Recent work by Houssel et al.~\cite{houssel_towards_2024} has shown that while LLMs may not be optimal for real-time threat prediction due to performance and computational constraints, they offer promising opportunities to enhance the interpretability of NIDS alerts. The same work has shown that LLM's explanations correctly retrieve information from the NetFlow data while being consistent with the feature values. Nonetheless, these models failed to reason logically and be factually accurate. \new{Overall, these LLMs can analyse a NetFlow sample as a whole and correctly identify traffic types (e.g., \textit{DNS} queries or \textit{HTTP} traffic). While that is useful for network operators, a simpler deterministic algorithm would achieve the same result.}
Furthermore, it struggles to correlate multiple features together to identify the nature of the attack correctly, instead, it treats individual features which appear suspicious to the model as independent explainability arguments for malicious activity. One significant problem is their tendency to hallucinate, by generating nonsensical or unfaithful content~\cite{minaee_large_2024}. These models lack comprehension of facts and logical reasoning, as it has been shown specifically for the network domain by Donadel et al.~\cite{donadel_can_2024}.

\new{This paper proposes a hybrid framework called \textit{eX-NIDS}, which is designed to complement existing NIDS. By augmenting LLM prompts with Cyber Threat Intelligence (CTI) and context‐specific knowledge, \textit{eX-NIDS} makes NIDS alerts explainable.} We evaluate this framework using the pre-trained models of Meta's \textit{LLama3}~\cite{touvron_llama_2023} and OpenAI's \textit{GPT-4}~\cite{achiam2023gpt} on a standardised NetFlow dataset, using a quantitative assessment of its potential to improve the explainability of NIDS. Our assessment framework evaluates the provided explanations for correctness, factual consistency and feature consistency.

\section{Related Work}
\label{sec:related-work}
Explainability is a crucial aspect of NIDS that enhances an operator's ability to respond effectively to threats. By integrating explainability with context-specific knowledge we can provide a clearer understanding of detected anomalies and facilitate a more informed response. Anomalies in network traffic, represented as NetFlows, are identified statistically by NIDS. However, certain attributes within NetFlows, such as an IP address with a known malicious reputation, might not directly impact the statistical anomaly detected but could indicate an underlying attack. LLMs present a promising solution to enhance the explainability of NIDS compared to existing ML explainability methods. LLMs can complement NIDS by tackling the lack of contextual insights or external knowledge existing explainable AI methods have. In modern NIDS, natural language data is largely absent, as these systems primarily analyse structured network telemetry. Most detection systems focus on network flows, aggregating packets sharing the same source and destination ports, IP addresses, and protocols into a single flow. This reduces memory complexity and simplifies analysis while preserving key semantics~\cite{catillo_machine_2023}. Despite the absence of natural language, several studies have explored using LLMs in NIDS to enhance explainability and detect malicious network flows~\cite{zhang_when_2024}.

While traditional LLM architectures are typically adapted to harness them for classification instead of natural language text generation, the sequence-to-sequence head is replaced by a classification head, matching the number of prediction classes, e.g., two for binary classification. 
\new{In most NIDS studies, LLMs are repurposed for classification rather than their original text‐generation role. To do this, researchers replace the sequence‐to‐sequence head with a classification head sized to the number of target classes (e.g. two for binary classification).}
Lai~\cite{lai_intrusion_2023}, Alkhatib et al.~\cite{alkhatib_can-bert_2022}, Ferrag et al.~\cite{ferrag_revolutionizing_2024}, and Lira et al.~\cite{g_lira_harnessing_2024} have investigated the use of LLMs like BERT~\cite{devlin_bert_2018} and GPT-3.5 \& 4~\cite{achiam2023gpt} for network intrusion detection, demonstrating the potential of transformer-based models. Their findings indicate that LLMs, when fine-tuned can detect attacks with acceptable accuracy, showcasing their adaptability. However, Houssel et al.~\cite{houssel_towards_2024} showed that without fine-tuning, models like GPT-4 and Llama3 perform poorly. In contrast, traditional ML models such as decision trees offer much faster inference times while being reliable at detecting threats, making them more practical for real-time NIDS. The inference time for LLMs is nearly 7,000 times slower than lightweight models, rendering them unsuitable for real-time applications. Manocchio et al.~\cite{manocchio_flowtransformer_2024} suggest that while LLMs may not be ideal for real-time detection, their foundational architecture, transformers, yield accuracy rates as high as existing state-of-the-art deep learning models. Transformers can identify network behaviours over a long term of time, which could be valuable in NIDS applications.

Several studies have explored existing explainable AI (XAI) methods to render NIDS more interpretable and explainable. These methods assign an importance score to each input feature, highlighting its influence on the model's output label~\cite{neupane_explainable_2022}. For example, Mallampati et al.~\cite{mallampati_enhancing_2024}, Senevirathna et al.~\cite{senevirathna_deceiving_2024}, and Zebin et al.~\cite{zebin_explainable_2022} apply SHapley Additive exPlanations (SHAP)~\cite{lundberg_unified_2017} across different network domains. SHAP quantifies the contribution of each feature, aiding real-time threat response by weighting features based on their impact on the model's predictions. Building on this, Wei et al.~\cite{wei_xnids_2023} propose a framework that uses feature importance scores to design dynamic intrusion response rules. Their approach identifies the features that drive prediction outcomes and generate defensive rules, providing insights into NIDS behaviour and facilitating troubleshooting when detection errors occur.

However, traditional XAI techniques like SHAP come with limitations. \new{Traditional XAI techniques like SHAP require users to have a strong background in machine learning to make sense of the results. They also only generate explanations based on statistical anomalies in the training data. In other words, they cannot incorporate contextual information that lies outside the original dataset.}\new{Finally these approaches operate within a vectorised feature space, assigning importance weights to each feature in relation to the model’s decision. While this can be informative for structured data, it is inadequate for a natural language input domain.}
Emerging studies suggest that LLMs could enhance NIDS explainability by generating human-readable, natural language explanations that go beyond statistical feature importance. Guastalla et al.~\cite{guastalla_application_2024} explored how LLMs can provide explanations for their predictions, marking an early foray into applying LLMs in this domain. Ali \& Kostakos~\cite{ali_huntgpt_2023} proposed \textit{HuntGPT}, which leverages GPT-3.5 to make decision-tree-based intrusion detection systems more explainable and interactive. Another study by Houssel et al.~\cite{houssel_towards_2024} highlighted that while LLMs struggle with detecting malicious NetFlows, they show promise in enhancing the explainability of NIDS. A conclusion shared by the previously mentioned studies as well. Nonetheless, these findings do not provide quantitative evaluations, leaving the potential of LLMs largely theoretical. A significant gap in this field is the lack of consensus on metrics for evaluating explainability in threat detection~\cite{neupane_explainable_2022}.

Ziem et al.~\cite{ziems_explaining_2023} tackles this issue by proposing an evaluation framework to assess LLM-generated explanations for decision-tree-based NIDS. Their novel approach involves using automatically generated quiz questions to measure human understanding of decision tree inferences. The study demonstrates that LLM-generated explanations significantly improve human comprehension of the model's classification decisions. Evaluators who received LLM-based explanations scored 39.5\% higher on quiz assessments compared to those given rule-based explanations. Moreover, human evaluators rated LLM explanations higher in terms of readability, overall quality, and incorporation of background knowledge. This work stands out as one of the first to provide a structured, quantitative evaluation of LLM explanations, emphasising their potential to enhance both NIDS functionality and threat response by improving user understanding of the system’s decisions.

The existing gap in the literature that our study addresses lies in the incomplete evaluation of explainability in hybrid detection systems using LLMs. Prior research by Ziem et al.~\cite{ziems_explaining_2023} evaluated LLM-generated explanations for NIDS by focusing on readability and participants' knowledge gained through quiz assessments, but did not rigorously evaluate the correctness of the explanations. Instead, the focus was on clarity and user comprehension without ensuring factual accuracy. The study also suggests room for enhancing explanation accuracy. Our study addresses this gap by proposing an approach to enrich LLM input prompts with contextual knowledge and CTI. Our work seeks to advance the practical application of explainability in NIDS using LLMs.

\begin{figure}[!t]
\centering
    \includegraphics[width=\textwidth]{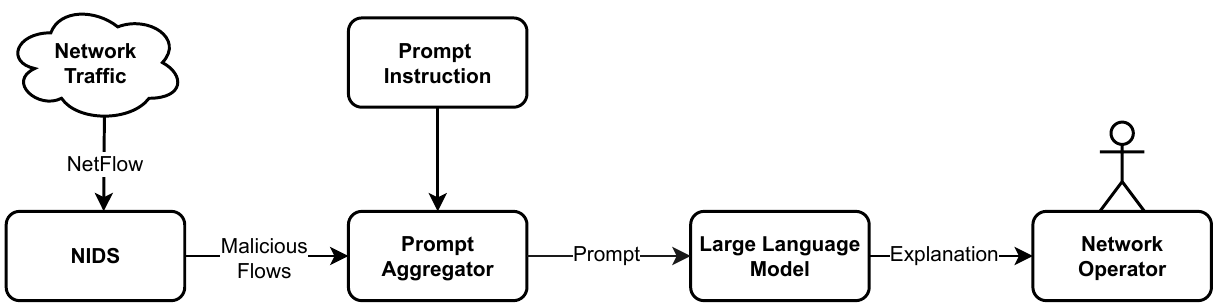}
    \caption{\textit{Basic-Prompt Explainer}.}
    \label{fig:flowchart-detailed-baseline}
\end{figure}

\section{eX-NIDS: A hybrid framework for explainable NIDS}
As discussed earlier in Section~\ref{sec:related-work} and shown in previous work~\cite{houssel_towards_2024}, LLMs have the potential to explain their classifications but struggle to efficiently identify malicious NetFlows. Hence, while LLMs may not be used as NIDSs, they can be employed in a hybrid solution, complementary to existing NIDSs, to enhance their explainability.

Building on this idea, we first establish a baseline framework that leverages LLMs to generate explanations for malicious flows identified by NIDSs. 
While this baseline model demonstrates the potential of LLMs in explaining NIDS decisions, it also produces many incorrect and misleading explanations.
To address this, we introduce a novel framework, \textit{eX-NIDS}, in the next subsection, which incorporates flow-specific contextual information to generate more accurate explanations.

\begin{figure}[!t]
\includegraphics[width=\textwidth]{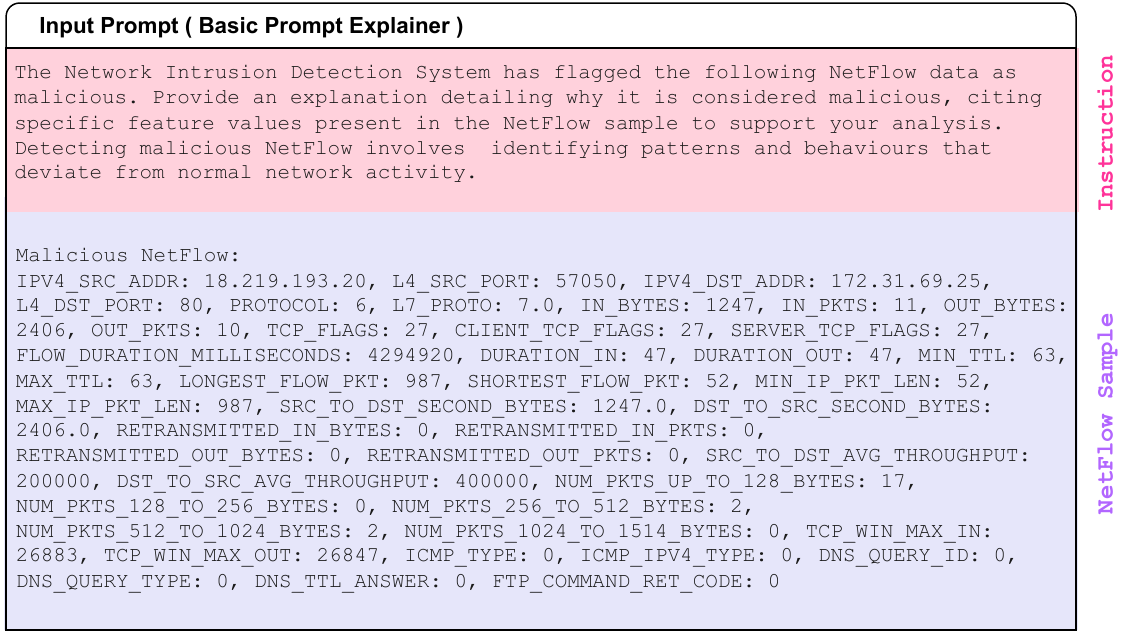}
\caption{Example input prompt provided to the LLM in the baseline framework. The first part (with a green background) contains the instructions, while the second part (with a blue background) presents the maliciously identified NetFlow sample.}
\label{fig:baseline-system-prompt}
\end{figure}

\begin{figure}[!t]
\includegraphics[width=\textwidth]{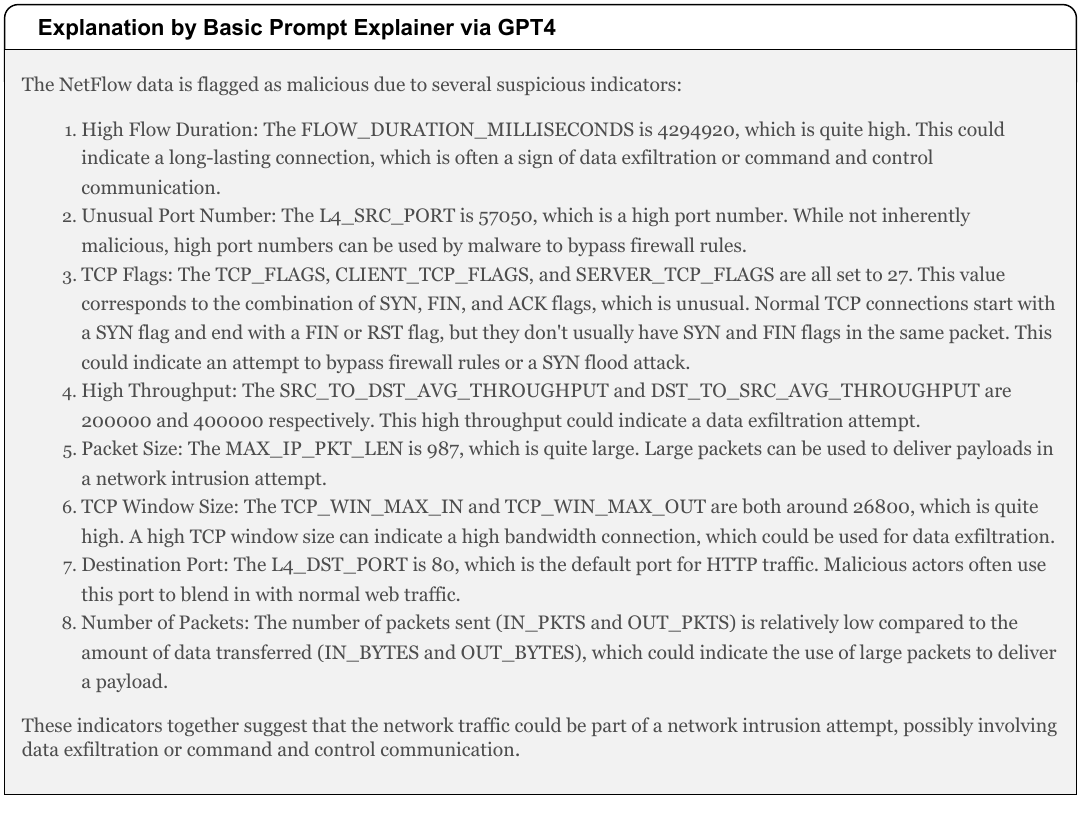}
\caption{Explanation provided by GPT4 for the prompt requested in Figure~\ref{fig:baseline-system-prompt} via the baseline framework.}
\label{fig:exp_basic_1}
\end{figure}

\subsection{\textit{Basic-Prompt Explainer}}
\label{sec:basicprompt}
Figure~\ref{fig:flowchart-detailed-baseline} shows our baseline framework.
In this framework, network traffic flows are passed to an existing ML-based NIDS in the form of NetFlow records that are then classified as either malicious or benign. 
The NetFlow records identified as malicious require explanation. Accordingly, they are sent for the explanation generation mechanism which starts with the prompt aggregator to be used to create an input prompt for the LLM, which will generate explanations for the malicious NetFlow records. 

In the prompt aggregator module, the given NetFlow is inserted into the input prompt template, which also includes input instructions.
The instruction piece of the input prompt provides clear indications to the LLM of the task to perform and the expected output. 
Figure~\ref{fig:baseline-system-prompt} shows an example input prompt generated by Prompt Aggregator in the baseline framework.
As can be seen, the LLM receives an input prompt that includes an instruction along with the maliciously identified NetFlow sample. 
Once the LLM receives the input prompt, generates an explanation in natural language that can be used by a human agent/network operator (Figure~\ref{fig:flowchart-detailed-baseline}).

\subsubsection{Explanations Generated by \textit{Basic-Prompt Explainer}}
Figure~\ref{fig:exp_basic_1} presents the explanations generated by the GPT-4 model for the prompt shown in Figure~\ref{fig:baseline-system-prompt}. A closer examination of these explanations reveals several inaccuracies. However, this issue is not unique to GPT-4. In fact, our evaluation of the explanations generated by the \textit{Basic-Prompt Explainer} using both GPT-4 and Llama3 indicated suboptimal correctness, with a significant presence of hallucinations that compromised reliability and usability.
Despite these shortcomings, the performance of the two models differed notably. For instance, GPT-4 accurately referenced present features almost every time, whereas Llama3 did so only about 80\% of the time.

By analysing the explanations provided by both models, we identified three distinct categories of incorrect explanations.\\
\textbf{- Hallucinations:} \\
\vspace{-2em}\begin{quote}
These are explanations which refer to feature values not present/different from input prompt. For instance, the two below examples:\\
\texttt{``The NetFlow data is considered malicious due to several suspicious feature values: MIN\_TTL:254, MAX\_TTL: 255 - These TTL values are unusually high, which could indicate a potential network attack.''}\\
\texttt{``Source IP and Port: The source IP address (172.31.69.17) and source port (50879) are not commonly associated with legitimate traffic.''}\\
\end{quote}\vspace{-1.5em}
\textbf{- Inconsistent reference to NetFlow features:} \\
\vspace{-2em}\begin{quote}
These are explanations that arise from a lack of knowledge of dataset features, similar to below example.\\
\texttt{``(Llama3) Packet length distribution: The presence of packets with lengths between 256 and 512 bytes (‘NUM\_PKTS\_256\_TO\_512\_BYTES') is unusual and may indicate an attempt to evade detection by packet inspection systems.''}\\
\end{quote}\vspace{-1.5em}
\textbf{- Misinterpretations of feature values:}\\
\vspace{-2em}\begin{quote}
As in below example in an explanation provided by Llama3, it is incorrectly claimed that port 54611 is unusual, even though it is a valid ephemeral source port chosen by TCP implementation software.\\
\texttt{``Unusual source port: The source port (54611) is not a commonly used port for legitimate traffic. It’s possible that an attacker is using this port to bypass firewall rules or evade detection.''}
\end{quote}

To mitigate these problems when utilising the LLMs to generate explanations for malicious NetFlow records, and to improve the accuracy, reliability, and domain relevance of these explanations, we propose a new approach in the next subsection. 
This approach leverages context-specific and CTI-augmented prompts to enhance performance and ensure more precise, trustworthy interpretations of network activity.

\begin{figure}[!t]
\centering
\includegraphics[width=\textwidth]{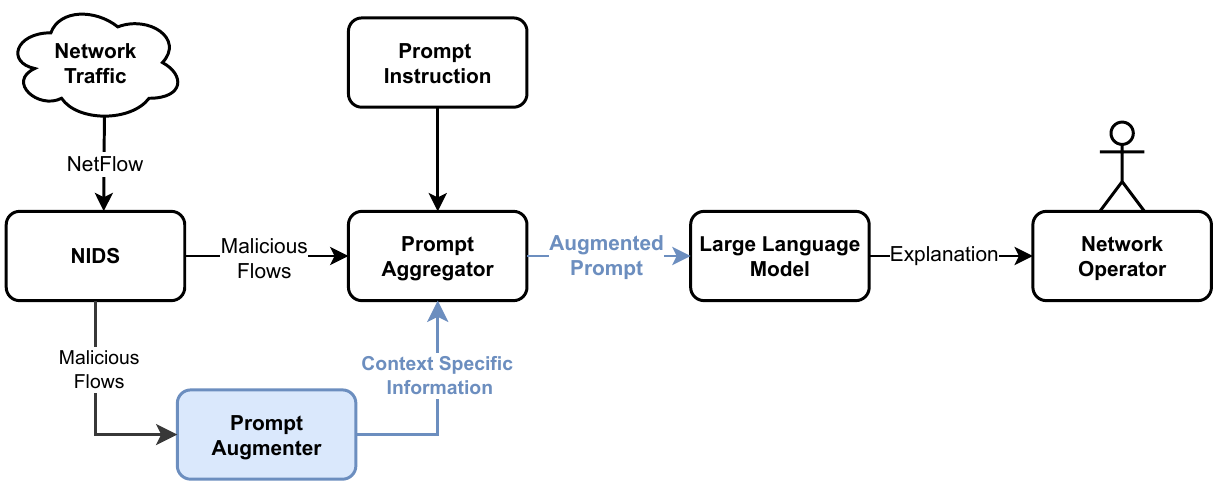}
\caption{High-level overview of eX-NIDS (proposed frameworks).}
\label{fig:flowchart-detailed-ex-nids}
\end{figure}
\begin{figure}[!b]
\centering
\includegraphics[width=\textwidth]{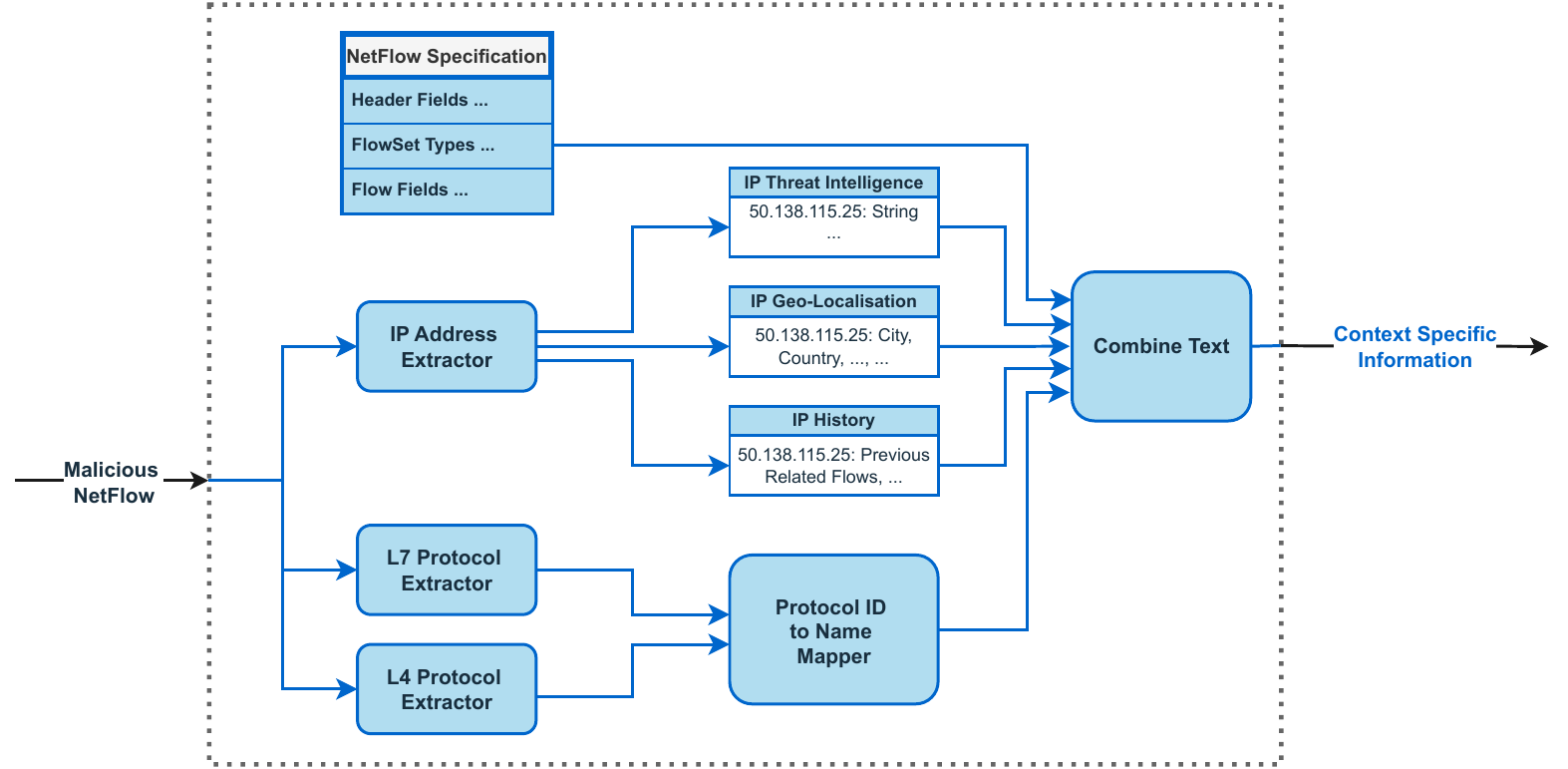}
    \caption{\textit{Prompt Augmenter} module. Connections to the external resources as well as the internal database are not shown here.}
    \label{fig:augmenter_module}
\end{figure}
\begin{figure}[!t]
\centering
\includegraphics[width=\textwidth]{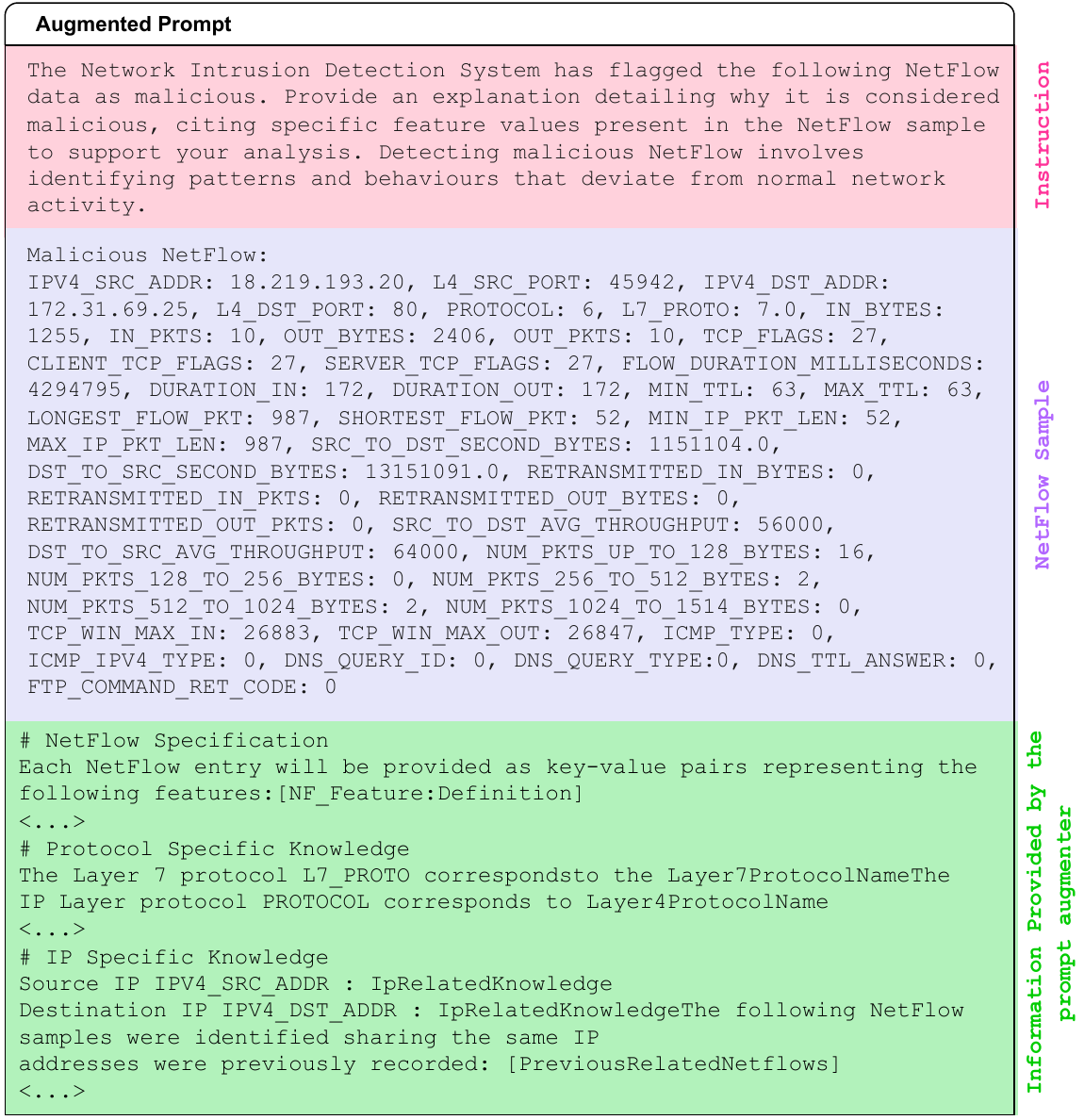}
    \caption{Augmented input prompt example used to request an explanation from the LLM for a detected malicious NetFlow entry, in the context of the \textit{eX-NIDS} framework.}
    \label{fig:ex-nids-explanation-prompt-template}
\end{figure}

\subsection{Augmented-Prompt Explainer (eX-NIDS)}
\label{sec:ex-nids}
Figure~\ref{fig:flowchart-detailed-ex-nids} presents the architecture of eX-NIDS, which extends the baseline framework by incorporating the \textit{Prompt Augmenter} to enrich the LLM's input prompt with contextual information.
Figure~\ref{fig:augmenter_module} details the \textit{Prompt Augmenter}'s architecture, which extracts and integrates three types of information: NetFlow specifications, context-specific knowledge related to IP addresses, and protocol identification numbers from the corresponding NetFlow sample.
Including the NetFlow specification—structured as feature name-definition pairs—helps prevent the LLM from misinterpreting feature meanings and measurement units.
Hallucinations often arise from errors in IP localisation, incorrect threat intelligence associations, or misinterpretations of protocol numeric IDs. To mitigate this, the Prompt Augmenter extracts IP Threat Intelligence, IP geolocation, and historical data on malicious NetFlows.
It queries an internal database using the extracted source and destination IP addresses to retrieve connection histories, while also fetching \new{geolocation} data and related threat intelligence.
Finally, based on the extracted Layer 7 (\textit{L7\_PROTO}) and Layer 4 (\textit{PROTOCOL}) identifiers, the module enriches the prompt with the corresponding protocol names (e.g. ID 17 maps to the User Datagram Protocol, UDP).

Figure~\ref{fig:ex-nids-explanation-prompt-template} presents an example of an augmented prompt. As illustrated, the information supplied by the prompt augmenter is appended to the end of the prompt generated by the \textit{Basic-Prompt Explainer} (Figure~\ref{fig:baseline-system-prompt}). The new section of the prompt, augmented contextual information, comprises three primary areas: \textit{NetFlow Specification}, \textit{Protocol Specific Knowledge}, and \textit{IP Specific Knowledge}, as previously explained. Due to space constraints, the complete details and the complete text of the example are not included here.

\begin{figure}[!b]
\includegraphics[width=\textwidth]{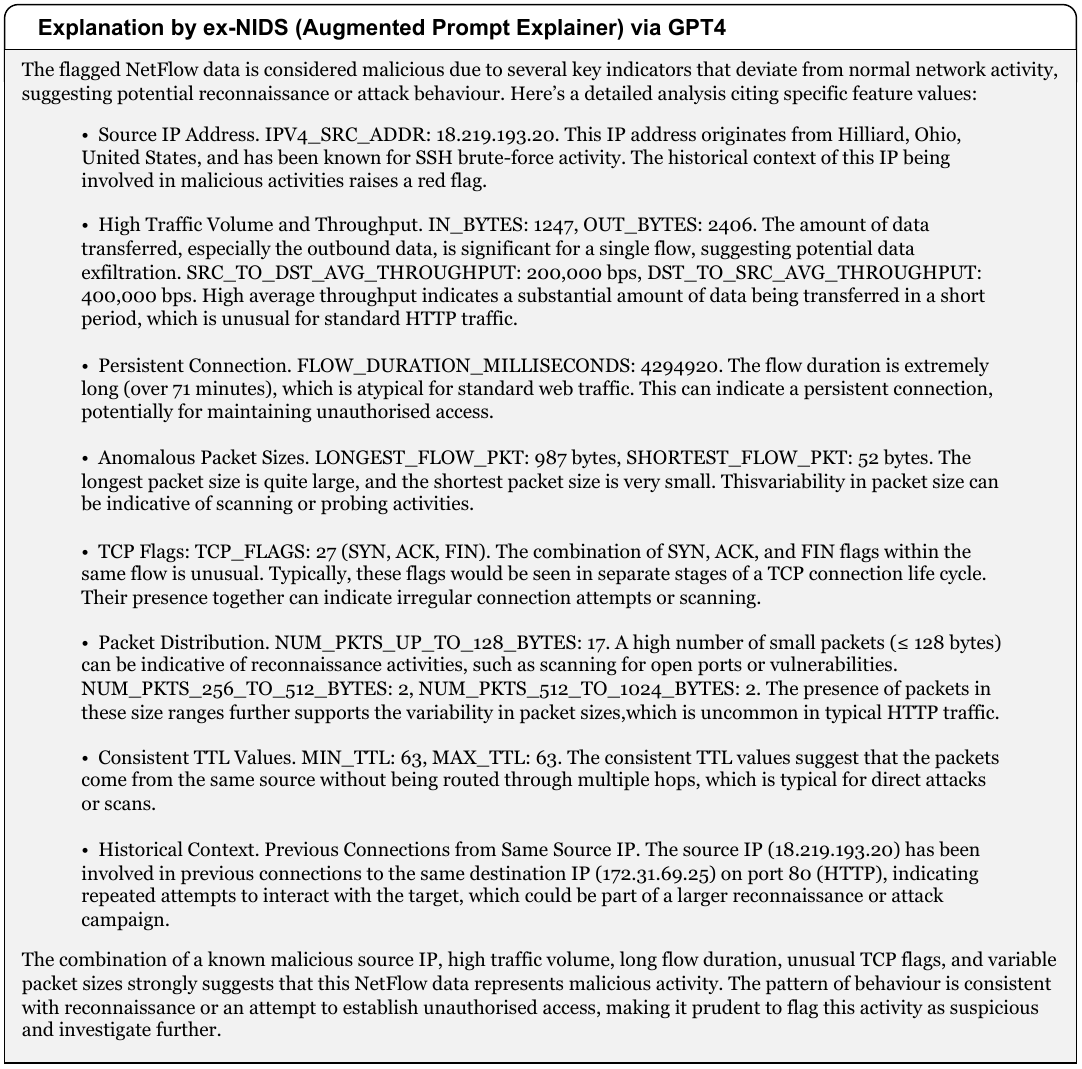}
\caption{Explanation provided by GPT4 for the augmented prompt requested in Figure~\ref{fig:ex-nids-explanation-prompt-template} via the ex-NIDS framework.}
\label{fig:exp_augmented}
\end{figure}

Retrieval Augmented Generation (RAG)~\cite{lewis_retrieval-augmented_2020} is not considered necessary in this context because NetFlow attributes are structured data, not natural language. Therefore, they can be efficiently searched using a traditional database without the need for embedding in a vector space. RAG enhances response accuracy by combining generative models with retrieval-based methods, integrating external information through a vector database. This database stores high-dimensional embeddings of data, enabling similarity searches to find relevant context for a query. This process allows the Language Model (LLM) to produce more factual and contextually rich explanations by grounding its outputs in external data. In our case, we ground the prompt in factual information retrieved from existing internal or external knowledge bases, without needing RAG's complex embedding process.

Figure~\ref{fig:exp_augmented} shows the explanation generated by ex-NIDS framework via GPT4 for the prompt shown in Figure~\ref{fig:ex-nids-explanation-prompt-template}. Comparing it with explanation generated by the \textit{Basic-Prompt Explainer} in Figure~\ref{fig:exp_basic_1} via GPT 4 (for the prompt shown in Figure~\ref{fig:baseline-system-prompt}) indicates that the inclusion of the contextual information significantly improves explanation accuracy. 
The explanations correctly reference feature values, using the appropriate units of measurement and demonstrating a proper understanding of their definitions. 
The explanation provided by eX-NIDS is indeed more effective because it provides a comprehensive analysis by considering multiple relevant factors and contextual information. In contrast, the baseline explanation (Figure~\ref{fig:exp_basic_1}) focuses on more trivial details like port numbers and lacks depth and context. For example, the baseline framework mentions trivial indicators which have no importance such as high port numbers and TCP flags while eX-NIDS identifies the context of importance and includes it in the explanation. Furthermore, it enriches the explanation by specifying the protocols and applying the correct units of measurement. 

\section{Explainability Evaluation Methodology}
Ex-NIDS demonstrates a significant improvement over the baseline explainer, providing more accurate and insightful justifications for its decisions. While there is still room for refinement, particularly in ensuring that all explanations correctly reflect the reasoning behind the model’s predictions, it represents a step forward in enhancing interpretability. However, some explanations do not always align with the actual malicious features of a given NetFlow sample. 

To systematically evaluate the quality and reliability of these explanations, we introduce a new method that quantitatively assesses their correctness, relevance, and consistency across different models. This approach enables a more rigorous comparison of explainability techniques, ensuring that they effectively highlight malicious behaviours within NetFlow data.
Accordingly, we equally sample 50 malicious NetFlows from the \textit{NF-CSE-CIC-IDS2018-v2}~\cite{sarhan2022towards} dataset. We only evaluate the explainability of the malicious flows, as these are the most relevant for industrial use cases. Each explanation provided is evaluated based on three metrics assessing the correctness and validity of the provided explanations.

\subsection{Evaluation Metrics}
All explanations are manually inspected by two domain experts according to the three following evaluation metrics which we define:

\begin{itemize}
    \item \textbf{Explanation Correctness}: This metric evaluates the \textit{semantic validity} of the explanation. The explanation should accurately interpret the features of the NetFlow data. For example, if an explanation argues that traffic is considered malicious due to high TTL values, even though it simply implies that the client's packets are being routed through a \new{few} hosts before arriving at the server, then the explanation would be considered incorrect. An explanation is correct only if all provided arguments are correct. The final metric is the percentage of correctly explained classifications for all explanations.
    \item \textbf{Feature Consistency}: This metric evaluates the model's ability to accurately use the feature values present in the input NetFlow data in its explanations. If the explanation includes these values but refers to them incorrectly or misunderstands the feature definitions, it is considered inconsistent. For example, if the model misunderstands the definition of average throughput and expresses it in bytes per second (Bps) instead of bits per second (bps), the explanation would be deemed feature-inconsistent. Similarly, if the model incorrectly indicates the presence of packets of a certain size when their value indicates their absence, it is considered to be feature-inconsistent. We consider an explanation feature consistent only if no feature inconsistency is present in the explanation. The final metric corresponds to the percentage of feature-consistent explanations.
    \item \textbf{Factual Consistency}: This metric evaluates the degree to which the explanation avoids non-factual or fabricated (hallucinated) information. An explanation is factually correct and grounded in the data, with no misinterpretations or invented details. For example, if the explanation accurately identifies that the BGP port number is 179 or correctly computes the TCP flags from decimal values, it would not be considered a hallucination. In contrast, explanations that introduce incorrect details, such as misinterpreting a feature’s value or fabricating non-existent features, are considered hallucinations. For instance, if the model states that 4294964 ms is equivalent to 43 minutes instead of the actual 71 minutes. The final metric corresponds to the percentage of factual consistent explanations over all explanations.
\end{itemize}

An inconsistent feature value or a hallucination does not necessarily imply that an explanation is incorrect. We evaluate each criterion independently, assuming the other criteria are correct. The evaluation results of our framework are compared against the baseline framework.

\subsection{Evaluated Models}
We define the specific experimental parameters tied to the evaluated LLM models to ensure the reproducibility of our results. 
In our study, we select the state-of-the-art LLMs as ranked by Chiang et al.~\cite{chiang_chatbot_2024}, using their \textit{ChatArena} benchmarking methodology. It is an open platform for evaluating LLMs based on human-evaluated preferences, it is maintained with an up-to-date leaderboard considering both open and closed-sourced models~\footnote{\href{https://chat.lmsys.org/?leaderboard}{LMSYS Chatbot Arena Leaderboard: chat.lmsys.org}}. During our study, LLama3 and GPT-4 emerged as leading contenders, excelling in areas like handling complex, domain-specific prompts in English and solving mathematical problems. More precisely, we select the instruction fine-tuned version of the LLama3 model because it can handle domain-specific tasks better.

To leverage the LLama3-70B-Instruct model~\cite{touvron_llama_2023} we employ the transformers python library. \new{Local inference was conducted on a Debian 12 system equipped with 32 GB of RAM, an NVIDIA GeForce RTX 3090 GPU, and an Intel i7-14700K CPU running at 3.4 GHz.} To perform inference on the GPT-4 model, we leverage the OpenAI API and their Python client, using the specific model \textit{gpt-4-1106-preview}~\cite{achiam2023gpt}. Our evaluation code is made available on GitHub~\footnote{\href{https://github.com/jetlime/eX-NIDS}{github.com/jetlime/eX-NIDS}}. We set the temperature to 0.7 and the maximum number of tokens to 2048, the default values across the technical implementations. In the context of LLMs, the parameter refers to the maximum number of tokens that the model can process in a single input or output, a requirement for long sequences of inputs and outputs we have in our context.

\section{Results}
We provide the experimental results obtained by our manual inspection in Table~\ref{tab:explainability-result-cse}. Compared to the baseline, our proposed prompt augmentation constantly increases the explainability quality over all 3 metrics and the two evaluated models. Most notably, it increases the factual consistency by 38\% and 14\% and the explanation correctness by 10\% and 40\%. We also notice a strong difference in performance between the two models, with GPT-4 scoring 20\% higher than LLama3. The weakest attribute remains overall the correctness of the explanations, which we observe among both models and the used prompts similar errors. These incorrect arguments argued in the explanations are due to the lack of reasoning and understanding of these models in the domain of networking has been shown by Donadel et al.~\cite{donadel_can_2024} by evaluating their understanding of network topology.

Overall, the GPT-4 model demonstrates superior performance, effectively employing the enriched prompts. While LLama3 continues to exhibit some hallucinations, these have been notably reduced and are now largely limited to minor errors, such as incorrectly converting time from milliseconds to seconds. This issue appears to stem from model-specific limitations, as GPT-4 does not exhibit the same error. Additionally, LLama3 shows lower correctness, as it fails to leverage information from the five previous connections associated with the given NetFlow, a context that GPT-4 successfully incorporates. This strong difference in performance, which is around 20\% can be explained by the fact that GPT-4 is most likely significantly larger, in terms of numbers of parameters, compared to the LLama3 model. Furthermore, GPT-4 has access to internal function calls to gain further access to knowledge bases~\cite{achiam2023gpt}.

\new{The LLM explanation layer introduces a notable performance overhead, which must be considered when deploying in real-time production environments. For local inference using the LLaMA 3 model, measured over 100 iterations, generating explanations took on average of 4 seconds for the \textit{Basic Prompt Explainer} and 6 seconds for the \textit{Augmented Prompt Explainer}. In contrast, inference via the OpenAI API incurs no computational cost but does involve a financial cost. In our dataset, basic prompts averaged 1,244 characters (461 tokens), while augmented prompts averaged 6,685 characters (2,308 tokens). At the time of experimentation, the OpenAI API pricing was \$2.50 per million input tokens and \$10.00 per million output tokens. Given that the average output length was 460 tokens for basic prompts and 562 tokens for augmented ones, the estimated cost per 1,000 queries was \$5.75 for the \textit{Basic Prompt Explainer} and \$10.37 for the \textit{Augmented Prompt Explainer}.}

\begin{table}[!t]
    \caption{Evaluation of our methods to render NIDS explainable on \textit{NF-CSE-CIC-IDS2018-v2}. The last column indicates the average over all three evaluation metrics.}
    \resizebox{\textwidth}{!}{
    \begin{tabular}{cccccc}
        \toprule
         \textbf{Model}&\textbf{Method}&\textbf{Explanation}& \textbf{Feature}&\textbf{Factual}& Avg.\\
         &&\textbf{Correctness (\%)}& \textbf{Consistency (\%)}&\textbf{Consistency (\%)}&Perf. (\%)\\
         \midrule
         LLama3&\textbf{Basic Prompt}&26 ($\pm$06)&84 ($\pm$05)&42 ($\pm$07)&50.66\\
         \cmidrule{2-6}
         70B-Instruct&\textbf{eX-NIDS}&\textbf{36} ($\pm$06)&\textbf{100} ($\pm$0)&\textbf{90} ($\pm$04)&75.33\\
         \midrule
         \midrule
        \multirow{2}{*}{GPT-4}&\textbf{Basic Prompt}&40 ($\pm06$)&96 ($\pm02$)&78 ($\pm$05)&71.33\\
         \cmidrule{2-6}
         &\textbf{eX-NIDS}&\textbf{80} ($\pm$05)&\textbf{100} ($\pm$0)&\textbf{92} ($\pm$03)&90.66\\
         \bottomrule
    \end{tabular}
    }
    \label{tab:explainability-result-cse}
\end{table}

\section{Conclusion}
Based on our findings, we advocate for the integration of LLMs as complementary tools within state-of-the-art NIDS frameworks. Their remarkable capacity to deliver enhanced explainability, particularly through detailed, contextually relevant explanations for triggered alerts, positions them as a powerful enhancement to traditional machine learning-based systems.
\new{However, simply feeding raw NetFlow data into an LLM and asking for an explanation leads to frequent hallucinations and misinterpretations, as we saw with the \textit{Basic-Prompt Explainer}. To avoid these pitfalls, a more structured approach, enriching prompts with contextual knowledge and CTI, is essential for accurate, dependable explanations.}

To address the initial challenges of hallucination and limited logical reasoning, we propose ex-NIDS framework that enhances the prompts with context-specific knowledge and CTI. This strategy not only effectively mitigates these shortcomings but also boosts the quality of explanations by over 20\%. Incorporating CTI and tailored contextual insights significantly elevates the explainability of LLMs, enabling them to produce outputs that align closely with network standards and NetFlow sample values. When paired with the right model, our framework delivers explanations that are consistently correct and factually grounded. Empirical experiments demonstrate that 80\% of explanations are accurate, 92\% remain factually consistent, and 100\% reliably reference the specific values of the NetFlow samples under consideration, marking a substantial advancement in the field.

Moving forward, future research should focus on evaluating the practical utility of these explanations through user studies involving network operators. Such studies would assess how different explainability methods impact operators' ability to mitigate attacks in real time, offering deeper insights into real-world usability. Although our current study focusses on evaluating the correctness of LLM explanations, this next step will be crucial in understanding their broader practical application. Furthermore, investigation and development are needed before they can be fully relied upon in critical security contexts. Future research should also prioritise improving LLMs' understanding of network traffic, an essential capability if they are to function as virtual system administrators supporting NIDS. Furthermore, upcoming studies should evaluate how LLM-generated explanations compare to those generated by traditional rule-based algorithms to provide actionable insights and mitigation strategies to network administrators.

\section{Acknowledgements}
This work does not raise any ethical issues and is supported by the University of Queensland School of Electrical Engineering and Computer Science (grant NS-2401).

\small
\bibliographystyle{unsrt}
\bibliography{main}
\end{document}